# Rendezvous Nodes (RN Nodes) Based Cluster Head Selection and Energy Efficient Data Aggregation with Mobile Sink for Lifetime Maximization in Wireless Sensor Networks.

Rajat Pushkarna
Nanyang Technological University, Singapore
rajat002@e.ntu.edu.sg

## ABSTRACT

Wireless sensor network (WSN) consists of a group of dedicated sensors nodes which are distributed over a certain area for observing and recording the physical conditions (like temperature, sound, pressure) of the environment and organizing the collected data at a central location. These Sensor nodes are powered by a battery, so energy conservation becomes an important aspect in WSN. The Paper proposes a way of combining traditional LEACH (Low Energy Adaptive Clustering Hierarchy) with Mobile Sink and Rendezvous Nodes (RN Nodes) along with some changes in terms of cluster head selection process. The paper aims at decreasing the number of dead nodes which in turn will relate to a increase in the total network lifetime. Particularly for large regions the proposed algorithms outperforms traditional Leach as proved in the Simulation Results.

## 1. INTRODUCTION

WSN is a network consisting of small and battery-powered sensor nodes deployed in the environment in order to sense and report, and also to collaborate in order to fulfill the overall system's goal. Wireless sensor network consists of hundred to even thousand sensor nodes. There exists a base station to which these nodes transmit the data. Since the energy of the node is finite and is supplied by a battery, any long distance communication will not be efficient and hence the network lifetime would be very less.

WSNs are used to measure environmental conditions like pollution levels, temperature, sound, humidity, wind speed and direction. The sensor node equipment has a radio transceiver and is also supported by an antenna, a microcontroller, an interfacing electronic circuit, and an energy source, usually a battery. Initially WSNs were used for military operations but its application has since been extended to health, traffic, and many other areas [1]. There exists a base station to which these nodes transmit data. Since the energy of the node is finite and is supplied by the battery, any long distance communication will not be efficient and hence the lifetime of the node will not be optimum. We propose a way of combining the traditional LEACH with a mobile sink and Rendezvous Nodes combined with changes in the cluster head selection process. The main idea behind this work is to decrease the number of dead nodes and thereby increase the total network lifetime.

## 2. RELATED WORKS

There have been a number of researches going on in the field of Wireless Sensor Networks. The basic aim is to increase the network lifetime by somehow decreasing the energy consumption inside a network.

Multi-Hop scenario is a good way to increase the network lifetime. In a multi-hop network each node transmits the data to other node and then data is send to the sink. This decreases the total distance and total energy consumption as other nodes will act as routers. Though, a drawback of using this approach would be that the nodes nearer to the sink would die faster.

One more approach is by going for clustering in the network. In this approach Cluster heads would be formed for each set of nodes and then these nodes will transmit information to Cluster Head and Cluster head will then transmit to Base station. This type of approach guarantees a sustained network lifetime. Although clustering will tend to reduce energy consumption, it has some problems. The main problem being the energy consumption concentrated more on the cluster heads, this issue in clustering, of how to distribute the energy consumption must hence be solved. LEACH [2] is a representative solution of this problem, which is a clustering method based on a probability model.

LEACH (Low Energy Adaptive Clustering Hierarchy) is a type of clustering algorithm where Cluster Head selection process occurs periodically. LEACH has basically 2 phases 1) Setup Phase and 2) Steady Phase. In Setup Phase Cluster Heads are formed based on a probability value and in Steady phase all the data transmission takes place. When clusters are formed, the nodes transmit their data through to the CH. LEACH uses time division multiple access (TDMA) schedule to decide when each node can transmit data to CH. In LEACH, the CHs are selected randomly and a node decides to which cluster it belongs. . The CH receives information from the nodes and decreases the total number of bits by aggregating them and transmitting the effective aggregated data to the sink. LEACH performs better than the other conventional protocols as the Cluster heads are formed randomly on a probability value and also the data aggregation is performed which in case decreases the energy consumption within the network and increases the network lifetime.

There are many improvements that have been proposed in the context of LEACH. LEACH-C, LEACH-I, V LEACH, TL LEACH, Multi-hop LEACH was formulated to improve the performance of the existing LEACH. Various other algorithms such as PEGASIS, TEEN, HEED, EACHP, Multi-hop Relay. Sometimes Cluster Head selection process is dependent on various factors which lead to even more sustained network lifetime of WSN.

Even Mobile Sinks [3] [8] [11] [13] [14] [19] [20] are also used sometimes to increase the network life time. Mobile Sink is basically a base station that will move around the sample region and collect information from the nodes. Its movement can be controlled or it may move randomly.

Since Mobile Sink cannot go to every node and collect data, concept of Rendezvous Nodes [4] [6] came into being. Rendezvous Nodes act as a temporary storage point and helps the nodes to communicate with Mobile Sink. As soon as Mobile Sink approaches the RN node a beacon signal is sent and RN node transfers the data to the Mobile Sink. Combined effect of RN node and Mobile Sink [5] is that it leads to more energy efficiency in case of WSN.

Not only a single sink, even the concept of multiple Mobile sink has been implanted by various researchers [9] [12] [18], but in this study we will take only the concept of a single Mobile Sink. Some studies have also elaborated about the delay in sending the information with use of Mobile Sink [10].

In the present paper concept of Mobile Sink and RN node has been combined with LEACH with changes in Cluster Head Selection process. The proposed method increases the life time of WSN.

## 3. PROPOSED ALGORITHM

The present study similarly has 2 stages as in traditional LEACH (Setup Phase, Steady Phase). The Setup phase includes the selection of Cluster head and RN nodes while the Steady State Phase deals with data transmission. The study considered a system composed of a fixed number of nodes placed randomly in a sampling region. All nodes have initial energy equal to $E_o$ which is taken to be 0.3J and a MS that travels through the middle of the sampling region. Mobile Sink is assumed to be having unlimited energy. It is also assumed that the locations of the nodes and sink can be found out for each round.

The proposed algorithm was split into several rounds that are similar to LEACH. Each node generates information with the same rate. The cluster head selection process depends on distance of node from the sink, residual energy value, number of times a node has been Cluster head and distance between other cluster heads. During simulation the traditional LEACH (with Static Sink) is compared with LEACH having Mobile Sink and RN nodes and even with our proposed LMRNACH (Leach Mobile Sink RN with altered Cluster Head Selection process).

### 1.1) Rendezvous Node

As already defined RN nodes act as a storage node and it passes the data to Mobile Sink. Several nodes usually have the prerequisites to be an RN. Initially each node is assumed to be a NN and all nodes decide whether or not they meet RN conditions. RN label is attached to the node which satisfies the required condition. The most important condition for an RN is distance from the MS as shown in Eq. (1):

if xm/2(1-Rthr) <= Nloc <= xm/2(1+Rthr)       (1)

where $x_m$ is the width of the sampling region, Nloc is the location of the node in the y-dimension and $R_{thr}$ is a constant with a value of <1.

### 1.2) Cluster Head Selection

The cluster head selection process depends on distance of node from the sink, residual energy value, number of times a node has been Cluster head and distance between cluster head and other nodes [7] [16] [17].

$$\begin{cases} Z(n) & \text{if } E(n) >= t1*(1/N)* \sum E(i) \quad \text{Tthresh} \\ 0 & \text{if } E(n) < t1*(1/N)* \sum E(i) \end{cases}$$

where Z(n)= a1T1(n) + a2T2(n) + a3T3(n) + a4T4(n). a1,a2,a3,a4 are 4 weighted constant parameters applied to adjust the relative influence of 4 sub-threshold terms T1(n),T2(n), T3(N) and T4(n).The sum of a1+a2+a3+a4 is always equal to 1 as to normalize the threshold value in contrast to traditional Leach.

$$T1(n) \begin{cases} \dfrac{P* E(n)}{(1/N)* \sum E(i)} & \text{if } E(n) >= t2*(1/N)* \sum E(i) \\ 0 & \text{if } E(n) < t2*(1/N)* \sum E(i) \end{cases}$$

$$T2(n) \begin{cases} \dfrac{P*(1/N)*\sum ds(i)}{ds(n)} & \text{if } ds(n) >= t3*(1/N)* \sum ds(i) \\ 0 & \text{if } ds(n) < t3*(1/N)* \sum ds(i) \end{cases}$$

$$T3(n) \begin{cases} \dfrac{P*(1/Q)*\sum dch(n,i)}{(1/Q)*(1/Q-1) \sum\sum dch(j,i)} & \text{if } Q>1 \\ (P*2*dch(n,Q))/M & \text{if } Q=1 \\ P & \text{if } Q=0 \end{cases}$$

$$T4(n) \begin{cases} \dfrac{P*(1/N)*\sum Nch(i)}{Nch(n)} & \text{if round}>1 \\ P & \text{if round}=1 \end{cases}$$

Here N is total number of alive nodes in current round, M is length of the square workspace(meter),Q is number of elected CH in current round, E(n) is current residual energy of node n (in Joule), P is desired percentage of nodes to be CH, n is node index(n=1,2,3….N), i is also node index (i= 1,2,3….N), j is CH index (j=1,2….M), ds(n) is distance from node to sink(meter), dch(j,i) is distance between node i and CH j(meter) and Nch(n) is number of rounds that node n was CH.

### 1.3) Overall Operation

During the Setup Phase every node is initially assumed to be Normal node. A node becomes a RN node if it satisfies equation 1. Every node then chooses a random number between 0 and 1.If the random number that is generated is less than the threshold value Tthresh then node is elected as CH. The CH and RN transmit their advertisement message to the Normal Nodes. Every node decides its cluster and which RN is nearest. Decision is based on minimum communication distance (dependent upon the strength of the signal of the advertisement message from the CHs and RNs).Based on the number of nodes in its cluster, a cluster head will create a TDMA schedule (so that nodes can identify when to send the data to CH) and it is broadcasted back to the non-CH nodes in its cluster. Once TDMA schedule is fixed, data transmission begins. In the steady phase nodes begin to transmit the data to CHs. When all data gets received, CH aggregates the data [15]. In LEACH transmission of data from CH to MS requires high consumption of energy. In our proposed algorithm CH will send data to MS if distance between them is less than the threshold distance do.

Where

$$d_o = \sqrt{E_{fs}/E_{amp}}$$

Efs is energy consumed by power amplifier in free space model and Eamp is energy consumed by power amplifier in multipath model.

If distance between CH and MS is greater than do(threshold distance), then CH transmits the data to nearest RN and then RN transmits to MS. This is done because if the transmission distance is more than 'do' then energy consumed is proportional to d^4 and if transmission distance is less than 'do' then energy consumed is proportional to d^2.

*1.4) Energy Model*

In order to calculate energy consumption first order radio communication model is used.

Energy consumed to transmit l bits of message to location d meters in distance is-

$$ETx = ETx\_elec(l) + ETx\_amp(l,d) \begin{cases} l*ETx + l*d^2 & \text{if } d \leq do \\ l*ETx + l*d^4 & \text{if } d > do \end{cases}$$

ETx is energy consumed by radio circuit for 1 bit transmission.

Total Energy consumed to transmit and receive data is

Etotal = ETx(l,d) + ERx(l).

where ERx is energy consumed by a node in receiving mode.

---

Xsink = 0 , Ysink = ym/2

Distribute n node randomly in the sampling region their initial energy set to S(i). Energy = Eo
  For i = 1:n
      S(i).type = "NN"  (All node assumed to be "NN" type)
  end for
  For r = 1 : r max   (start Round, r max is final round)
      For i = 1:n
            If |Y(i) – ym /2| < Rthresh
                  S(i).type = "RN"
            End if

            If S(i).Energy <= 0
                  S(i).type = "dead"
                  S_dead(r) = S_dead (r) +1
            else

                Remain. Eng total(r) = ∑E(n)
            End if
              If S(i) is first_dead
                      Flag_First_dead(r) = 1
              End if

              If S(i) is 25% dead node
                    Flag_25%_dead (r) = 1
              End if
      End for
  Avg – Eng – Node (r) = Remain. Eng total(r) / (n – n dead(r))
For i = 1: n

CH Selection Process if using traditional leach use probability formula, for proposed algorithm use Tthresh.

          If ( S(i).type ~= "RN"
      (depends on algorithm scenario )

              If ( S(i).Energy => Eaveragenet (r)
          (depends on algorithm scenario)

                S(i).type = "CH"
              end if

        end if

    end for

for i = 1: n

      If S(i).type == "NN"

Find nearest "CH" and transmit normal node data to nearest CH

    S(i).Energy = S(i).Energy - ET  and  reduce receive energy from selected CHs initial energy.

      End if

  If S(i).type == "CH"

    If dist_from_CH_to_sink(i) > do

        Find nearest lived RN

        Transmit data to selected RN and reduce transmit energy from its Initial energy

          If |YRn - Ysink| < do

```
Transmit data to selected RN and reduce transmit energy from RN's
Initial energy

        End if

    End if

  Else
Transmit data from CH to sink directly

        S(i)ChEnergy = S(i)ChEnergy-Er
    End if

    End for

Xsink= Xsink +ΔX
Ysink= Ysink +ΔY

End for
```

Pseudo Code of Algorithm

## 4. SIMULATION

For the simulation of traditional LEACH with Mobile Sink and RN node five scenarios were taken into account.

1) Mobile Sink-1 – In this Scenario RN's do not participate in the CH selection process and all other nodes can be CH based upon threshold.

2) Mobile Sink-2 – In this Scenario RN's do not participate in the CH selection process and all other nodes can be CH based upon threshold energy, the remaining energy should be greater than the average energy.

3) Mobile Sink-3 – In this Scenario RN's participate in the CH selection process and all other nodes can be CH based upon threshold.

4) Mobile Sink-4 – In this Scenario RN's participate in the CH selection process and all other nodes can be CH based upon threshold but there energy remaining should be greater than the average energy.

5) Static Sink- In this Scenario our Sink does not move and is considered to be positioned at the centre of the sampling region. (Traditional LEACH case).

For the proposed algorithm, in which cluster head selection process will depend on four factors, 2 scenarios were taken into account namely Proposed Mobile Sink 2(PMS2) and Proposed Mobile Sink 4(PMS4). MS1 and MS3 are not considered since energy value is taken into consideration during the formulation of Tthresh. Simulation of these scenarios was done for 4 different regions 200m, 250m, 300m and 450m.

Simulation Parameters

| Parameters | Value | Description |
|---|---|---|
| Xm, Ym | 200m,250m,350m,450m | Length and Width of Region |
| Eo | 0.3J | Initial Energy |
| ETx | 50 nJ/bit | Energy consumed by radio in transmit mode |
| ERx | 50nJ/bit | Energy consumed by radio in receive mode |
| EDA | 5nJ/bit/signal | Energy consumed for data aggregation |
| Eamp | 0.0013pJ/bit/m^4 | Energy consumed by amplifier in multipath |
| Efs | 10pJ/bit/m^2 | Energy consumed by amplifier in free space |
| l | 4000bit | Data packet Size |
| Rthresh | 16% | Constant related to width of sampling region |
| n | 100 | Number of nodes |

**Traditional Leach with Mobile Sink and RN node**

Table-1-First Dead Node (in rounds)

| Dimension (in m) | Mobile sink-1 | Mobile sink-2 | Mobile sink-3 | Mobile sink-4 | STATIC |
|---|---|---|---|---|---|
| 200*200 | 243 | 206 | 161 | 180 | 229 |
| 250*250 | 52 | 123 | 99 | 82 | 67 |
| 350*350 | 28 | 13 | 14 | 19 | 19 |
| 450*450 | 10 | 12 | 4 | 13 | 11 |

Table-2-25% Dead Nodes (in rounds)

| Dimension (in m) | Mobile sink-1 | Mobile sink-2 | Mobile sink-3 | Mobile sink-4 | STATIC |
|---|---|---|---|---|---|
| 200*200 | 553 | 566 | 625 | 648 | 487 |
| 250*250 | 418 | 312 | 545 | 539 | 300 |
| 350*350 | 220 | 161 | 257 | 278 | 110 |
| 450*450 | 80 | 93 | 102 | 87 | 40 |

Table 1 shows the occurrence of first dead node (in rounds) in traditional LEACH with Mobile sink and RN node.

Table 2 shows the occurrence of 25% dead nodes (in rounds) in traditional LEACH with MS and RN. These parameters are important as they help in assessing the quality of network and also the network lifetime (especially 25% dead nodes).As the dimension increases the performance gets better. The Mobile Sink algorithms outperformed the Static sink by long way as the dimension increases.

**Graph-1-** Number of dead nodes in Different Scenarios (450*450)

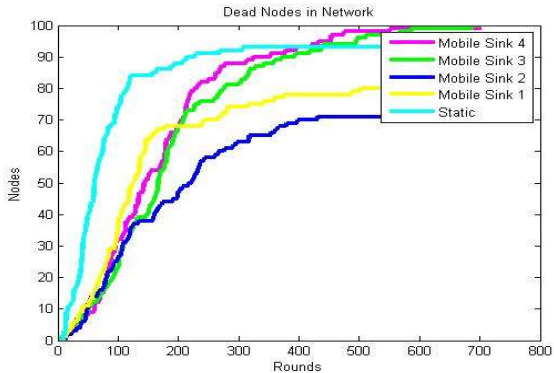

**Graph-2-** Number of alive nodes in Different Scenarios (450*450)

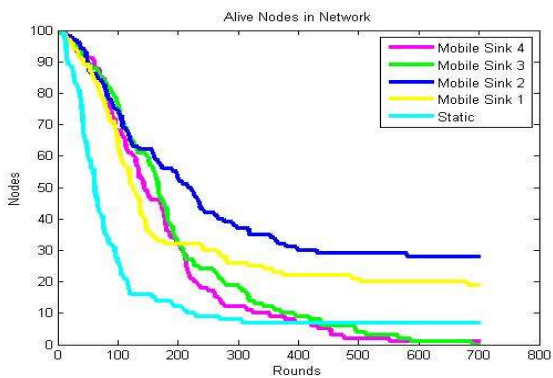

Graph 1 and 2 shows that how Mobile Sinks performs better than Static Sink in terms of network lifetime. By comparing the slopes it can be easily seen that in Static Sink nodes die faster and hence network lifetime decreases

**Proposed Algorithm Results**

Table-3-First Dead Node (in rounds)

| Dimension (in m) | Mobile sink-2 | Mobile Sink-4 |
|---|---|---|
| 200*200 | 178 | 88 |
| 250*250 | 29 | 43 |
| 350*350 | 4 | 4 |
| 450*450 | 2 | 2 |

Table-4-25% Dead Nodes (in rounds)

| Dimension (in m) | Mobile sink-2 | Mobile Sink-4 |
|---|---|---|
| 200*200 | 493 | 489 |
| 250*250 | 592 | 408 |
| 350*350 | 318 | 309 |
| 450*450 | 216 | 197 |

Table 3 shows the occurrence of first dead node (in rounds) in our proposed algorithm LMRNACH. Only 2 cases have been considered as the nodes are energy aware because of the threshold (Tthresh).

Table 4 shows the occurrence of 25% dead nodes (in rounds) in our proposed algorithm LMRNACH. When our Proposed Mobile Sink is compared with traditional LEACH the results were extremely good. As the network dimension increases the proposed LMRNACH outperforms the traditional LEACH with Mobile Sink. In network dimension of 450*450 proposed algorithm performed nearly twice as good as existing traditional LEACH with MS.

**Graph-3-** Number of dead nodes in Proposed Mobile sink (450*450)

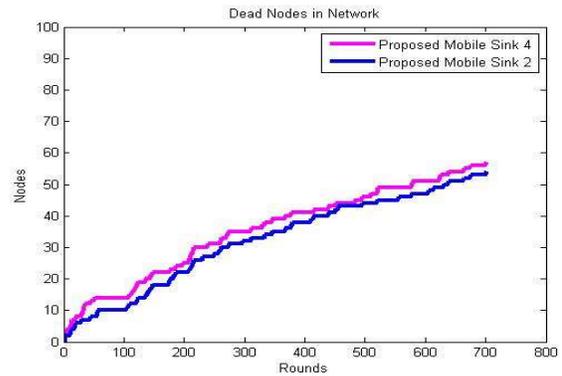

**Graph-4-** Number of alive nodes in Proposed Mobile sink (450*450)

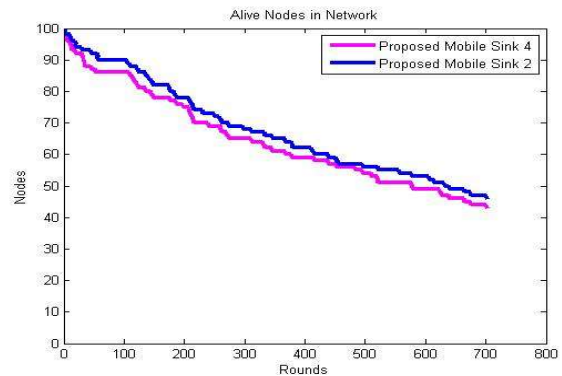

Graph 3 and 4 shows that PMS2 performs better than PMS4 this is due to fact that MS2 the RN nodes don't participate in CH election process and energy is not wasted in it, rather RN nodes simply pass the information directly to Mobile Sink.

**Comparison between proposed and existing**

**Graph-5-** Comparison between dead nodes in MS2 and PMS2 (450*450)

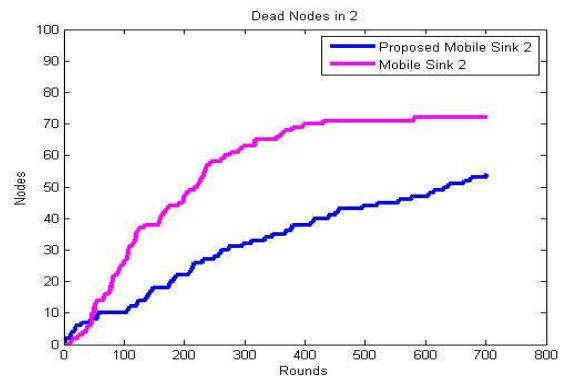

**Graph-6-** Comparison between alive nodes in MS2 and PMS2 (450*450)

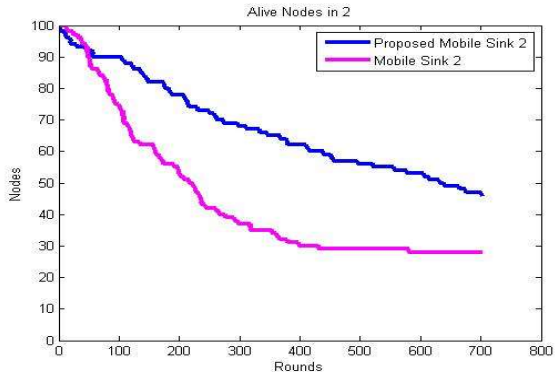

Graph 6 shows that in comparison with Mobile Sink 2 our proposed mobile sinkperforms way better. Numbers of alive nodes in Proposed Mobile Sink 2(PMS2) are greater than Mobile Sink(MS2).

**Graph-7-** Comparison between dead nodes in MS4 and PMS4 (450*450)

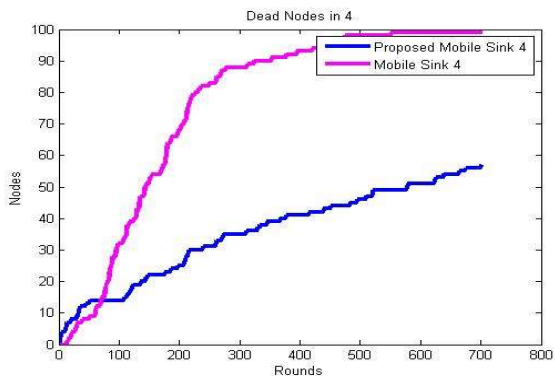

It can be inferred from graph 5 that Proposed Mobile Sink has less number of dead nodes as compared to Mobile Sink with traditional Leach. So it is clear from the simulation results that LEACH with Mobile Sink and RN nodes with altered Cluster Head Selection process (LMRNACH) is a good method to increase the network lifetime as compared to traditional LEACH with Static sink or even with Mobile Sink.

**Graph-8-** Comparison between alive nodes in MS4 and PMS4 (450*450)

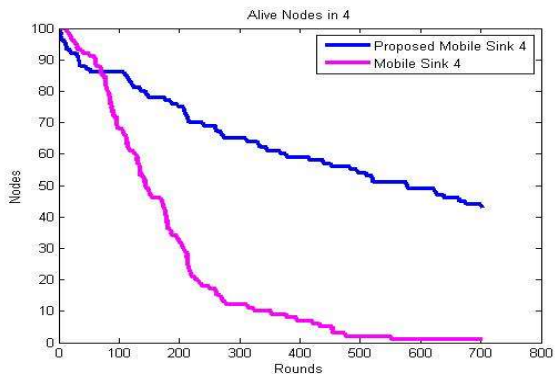

Graph 7 and 8 in a similar way shows that in comparison with Mobile Sink 4, PMS4 performs better and has a larger network lifetime.

**Graph-9-** Residual energy in all the scenarios(450*450)

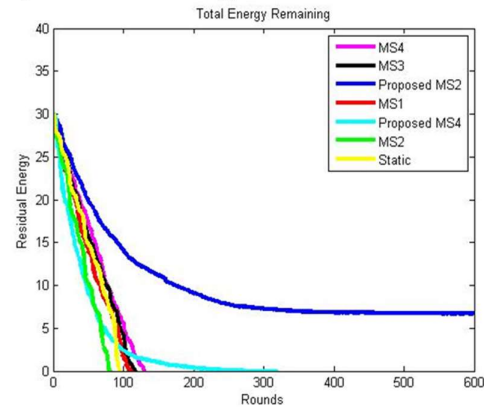

**Graph-10-** Alive nodes in the network in all scenarios (450*450m$^2$)

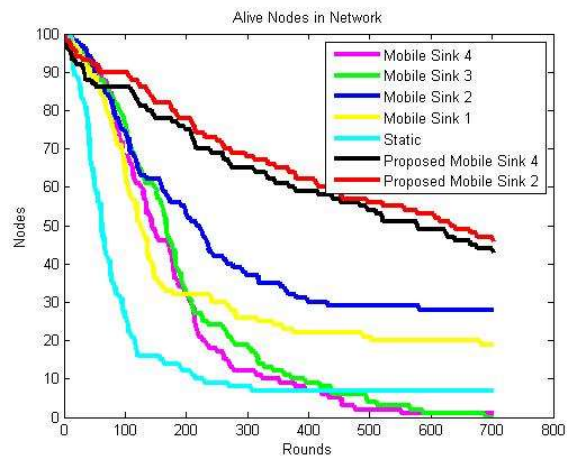

## 5. CONCLUSION

Network Lifetime is a key issue for any Wireless Sensor Network. The present study was aimed at prolonging the Network lifetime. The LMRNACH algorithm considered LEACH with Mobile Sink and RN nodes, with altered Cluster Head selection process. Even the simulation of traditional LEACH with Mobile Sink and RN node was performed. It was seen that in traditional LEACH case the Mobile Sinks (MS1, MS2, MS3, and MS4) performed better than the static sink.

LMRNACH was performed by altering the cluster head selection process and it was dependent on 4 factors distance of node from sink, distance of node from other CH, Residual Energy and Number of times a node has been CH. Simulation results clearly proves the proposed algorithm outperformed traditional LEACH mobile sink and RN node (particularly in large network dimension).

## 6. REFERENCES

[1] CHEE-YEE CHONG, Sensor Networks: Evolution, Opportunities, and Challenges IEEE 2003.

[2] Heizelman WR, Chandrakasan A, Balakrishnan H. Energy-efficient communica-tion protocol for wireless micro sensor networks. In: IEEE Hawaii international conference on system sciences. 2000. p. 10–20.

[3] Liu W, Lu K, Wang J, Xing G, Huang L. Performance analysis of wireless sensor networks with mobile sinks. IEEE 2012;61(July (6)):2777–88.

[4] Konstantopoulos C, Pantziou G, Gavalas D, Mpitziopoulos A, Mamalis B. A rendezvous-based approach enabling energy-efficient sensory data collec-tion with mobile sinks. IEEE Trans Parallel Distrib Syst 2012;23(May (5)): 809–17.

[5] Mohammad Reza Zahabi. Optimizing LEACH with Mobile Sink and RN nodes.2014(18 Aug). Science Direct.

[6] Maximizing the Lifetime of Wireless Sensor Networks Using Multiple Sets of Rendezvous 2015

[7] A new evolutionary based application specific routing protocol for clustered wireless sensor networks Science Direct 2015

[8] Gowri.K #1, Dr.M.K.Chandrasekaran M.E.,M.B.A.,Ph.D *2, Kousalya.K #3 A Survey on Energy Conservation for Mobile-Sink in WSN

[9] Andrew Wichmann Smooth Path Constructionand adjustment for multiple mobile sink in WSN 2015

[10] ESWC: Efficient Scheduling for the Mobile Sink in Wireless Sensor Networks with Delay Constraint IEEE 2013

[11] L.Brindha1 U.Muthaiah2 Energy Efficient Mobile Sink Path Selection Using a Cluster Based Approach in WSNs 2015

[12] Andréa Cynthia Santos, Christophe Duhamel b , Lorena Silva Belisário b Heuristics for designing multi-sink clustered WSN topologies 2016

[13] Yasmine Derdour The Impact of the Mobile Element on Performance improvement in WSN 2014

[14] Massimo Vecchio improving area coverage of wireless sensor networks via controllable mobile nodes: A greedy approach 2015

[15] Parabhudutta Mohanty Energy Efficcient Data Aggregation and delivery in WSN 2015

[16] S.H Kang Distance Based Thresholds for Cluster Head Selection in Wireless Sensor Networks IEEE 2012

[17] Saadat M Improving threshold assignment for Cluster head selection in WSN IEEE 2010

[18] David Jea  Arun Somasundara Mani Srivastava  Multiple Controlled  Mobile Elements  for Data Collection in Sensor Networks

[19] Lambrou Tp Exploiting Mobility for efficient coverage in WSN 2010

[20] Di Francesco M. Data Collection in Wireless Sensor Network with Mobile elements 2011